\documentclass[12pt]{iopart}

\expandafter\let\csname equation*\endcsname\relax
\expandafter\let\csname endequation*\endcsname\relax

\usepackage{graphicx,subfigure,amsmath,bbm}
\usepackage{graphicx}
\usepackage{dcolumn}
\usepackage{bm}
\usepackage{amsmath}
\usepackage{amsthm}

\usepackage[english]{babel}
\usepackage[utf8]{inputenc}
\usepackage[colorinlistoftodos]{todonotes}
\usepackage{bbold }
\usepackage{physics}
\usepackage[utf8]{inputenc}
\usepackage{amssymb }
\DeclareGraphicsExtensions{.eps}
\usepackage{epstopdf}
\usepackage{latexsym}
\usepackage{amsfonts}
\usepackage{xcolor}
\usepackage{units}
\usepackage{bm}
\usepackage{verbatim}
\usepackage[colorlinks = true,
            linkcolor = red,
            urlcolor  = blue,
            citecolor = magenta,
            anchorcolor = red]{hyperref}
\usepackage{esint}
\usepackage{soul}
\usepackage{cancel}
\usepackage[normalem]{ulem}
\usepackage{ dsfont }
\usepackage{empheq}
\usepackage{ wasysym }
\usepackage{scalerel}
\usepackage{soul}

\usepackage[]{ subfigure }

\DeclareMathOperator{\arccosh}{arcCosh}

\begin{document}

\title{Hybrid normal-superconducting Aharonov–Bohm quantum thermal device}

\author{Gianmichele Blasi}
\address{Department of Applied Physics, Université de Genève, 1211 Geneva, Switzerland}
\author{Francesco Giazotto}
\address{NEST, Instituto Nanoscienze-CNR and Scuola Normale Superiore, Piazza San Silvestro 12, 56127 Pisa, Italy}
\author{G\'eraldine Haack}
\address{Department of Applied Physics, Université de Genève, 1211 Geneva, Switzerland}

\vspace{10pt}


\vspace{10pt}

\begin{abstract} We propose and theoretically investigate the  behavior of a ballistic Aharonov-Bohm (AB) ring when embedded in a N-S two-terminal setup, consisting of a normal metal (N) and superconducting (S) leads. 
This device is based on available current technologie and we show in this work that it constitutes a promising hybrid quantum thermal device, as quantum heat engine and quantum thermal rectifier. Remarkably, we evidence the  interplay of single-particle quantum interferences in the AB ring and of the superconducting properties of the structure to achieve the hybrid operating mode for this quantum device. Its efficiency as a quantum heat engine reaches $55\%$ of the Carnot efficiency, and we predict thermal rectification factor attaining $350\%$.  These predictions make this device highly promising for future phase-coherent caloritronic nanodevices.
\end{abstract}

\maketitle

\section{Introduction}

Engineering versatile and efficient quantum thermal devices to manage heat at the nanoscale is highly challenging and relevant for future quantum technologies \cite{fornieri2017towards,pekola2021colloquium,giazotto2006opportunities}. Controlling heat implies, in particular, being able to exploit a temperature gradient as a resource for operating a device as heat engine \cite{josefsson2018quantum,ono2020analog,germanese2022bipolar,scharf2020topological,marchegiani2020superconducting}, and to allow for preferential heat flow in one direction under thermal biasing, a feature known as thermal rectification \cite{ fornieri2017towards, Senior2020, martinez2015rectification}. Both abilities have been investigated independently from each other, motivating numerous proposals and experiments that exploit various platforms with the objective of managing heat at the nanoscale in an efficient way. As recent examples, we emphasize achievements with superconducting circuit QED setups \cite{Saira2007, Ronzani2018, Senior2020},  with superconducting-semiconducting devices \cite{Fornieri2014,Fornieri2015, Iorio2021} and with graphene-based samples \cite{Wang2017} for heat rectification. The first experimental heat engines at the nanoscale have exploited among others features trapped ions \cite{Ronagel2016, Chand2017} and cold atoms \cite{Brantut2013}, NV-centers samples \cite{Klatzow2019}, semiconducting quantum dots \cite{josefsson2018quantum} and, very recently, superconducting tunnel junctions \cite{germanese2022bipolar}. \\

\begin{figure}[t]
	\centering
	\includegraphics[width=0.9\linewidth]{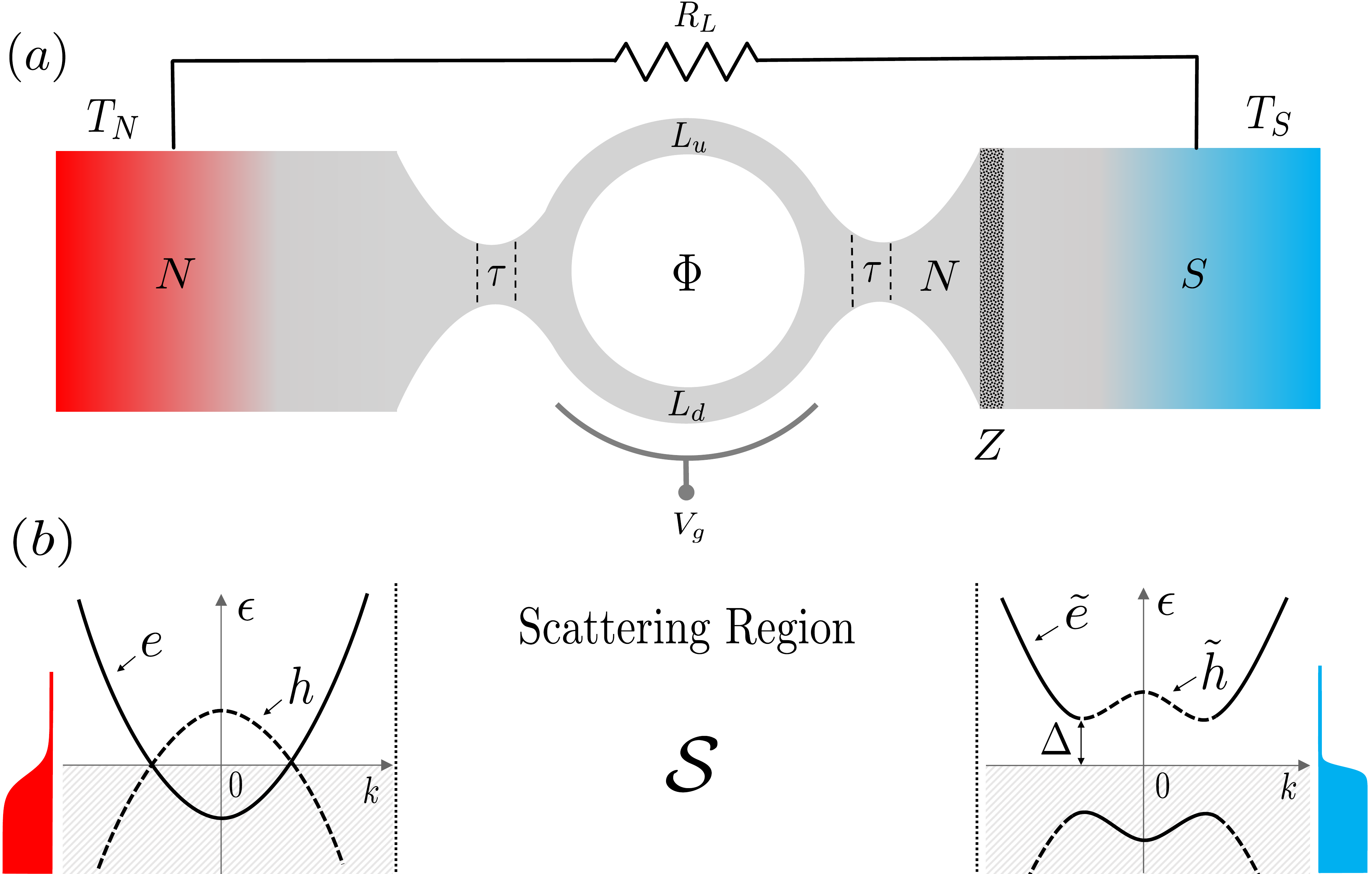}
 	\caption{Two-terminal N-S AB quantum thermal device. (a) Scheme of the central AB ring with magnetic flux $\Phi$, gate voltage $V_g$ and length imbalance $\Delta L=L_u - L_d$ as tunable parameters. The AB ring is connected to the contacts through two T-junctions with transmission probability $\tau$, and through a N-S junction with the superconducting contact. The N and S contacts are biased in temperature, $T_N - T_S = \Delta T$. The circuit can be closed with dissipationless (superconducting) wires with a load resistance $R_L$ to operate the device as quantum heat engine. (b) Dispersion curves for electron-like quasiparticles (solid curves) and hole-like quasiparticles (dashed curves) respectively in the normal lead (left) and superconducting lead (right). The presence of the superconductor opens a gap $\Delta$ in the eigenspectrum, and hybridizes electron- and hole-states. $\mathcal{S}$ indicates the scattering matrix describing the central scattering region accounting for the presence of the AB ring, the T-junctions and the N-S interface.}
	\label{System_main}
\end{figure}

Towards the development of realistic and  efficient quantum technologies, it becomes clear that hybrid quantum thermal devices that can combine several tasks towards heat management at the nanoscale become highly desirable. Their characterization have been the topic of recent theoretical investigations \cite{Manzano2020, Ito2018, Yang2015,Sanchez2016a}. Here, we propose and theoretically investigate an Aharonov-Bohm (AB) ring embedded in a normal-superconducting (N-S) two-terminal device as a versatile and efficient quantum thermal machine. We demonstrate that both single-quantum interferences and superconducting contact allow for operating this device as a quantum heat engine and a quantum heat rectifier. This configuration is inspired by recent works demonstrating on the one side, excellent thermoelectric response of an AB ring in a normal two-terminal setup \cite{Haack2019, Haack2021} and on the other side, high rectification coefficients by exploiting a left-right asymmetry in the density of states of the contacts induced by the superconducting gap compared to the normal metal \cite{fornieri2017towards}. From a theoretical point of view, we characterize the behavior of this N-S AB interferometer within and beyond the Andreev approximation, evaluating the functionning of N-S structures as quantum thermal devices in full generality. \\


\section{Theoretical framework}

We analyze the thermodynamic performances of a Aharonov-Bohm ring in a N-S two-terminal setup within a scattering-matrix formalism~\cite{Lesovik2011a}. This theoretical approach is motivated by its ability to keep track of the quasiparticles phase coherence in the device and of the specificities of the superconducting contact characterized by its gap $\Delta$ and its critical temperature $T_C$. In contrast to previous works \cite{Haack2019, Haack2021} investigating the AB ring as efficient thermoelectric device when connected to two normal contacts, we account for the electron-like and hole-like behaviours of the quasiparticles being subject to a temperature bias between the two contacts, one normal metal contact and one superconducting contact. It is the combination of these N-S contacts and the AB ring that allows this device to become hybrid -- it acts as heat rectifier and heat engine with large efficiency in both operating modes. 

\subsection{Model for a N-S AB ring}

The device is sketched in Fig.~\ref{System_main}: it is made of an AB ring as central part, connected to the left and right contacts, a normal contact ($N$) and a superconducting contact ($S$). The two contacts are described through their chemical potential $\mu_{N,S}$ and temperature $T_{N,S}$. The AB ring encloses a magnetic flux $\Phi$, and is characterized by the length of its upper and lower arms, respectively $L_u$ and $L_d$. A top gate allows for a gate voltage $V_g$. The AB ring is connected to the two leads through T-junctions characterized by their transmission probability $\tau\in [0,1]$ \cite{Buttiker1984}.  
The superconducting contact is connected to the central part through a N-S (Normal-Superconductor) junction.
We model this junction with a delta-like contact resistance at the interface between the normal and superconducting parts, following Refs.~\cite{BTK1982,Blasi2021}. This barrier is characterized by a dimensionless parameter $Z$: for $Z \ll 1$, the interface is said to be very transparent and ideal for $Z=0$, whereas $Z \gg 1$ corresponds to the tunnel limit. 

Within a scattering matrix approach, we evaluate the charge and heat currents in the normal (left) contact, denoted as $I_N$ and $J_N$ respectively. They can be expressed in the compact form \cite{Lambert1998a,Benenti2017a,Blanter2000,Blasi2020}:
\begin{equation}
\label{current_Lambert_general}
\begin{bmatrix}
I_N \\
J_N 
\end{bmatrix}
 =\frac{2}{h}\!\!\sum_{\substack{j= N,S \\\alpha,\beta = \pm}}\!\int_{0}^{\infty} \, d\epsilon 
\begin{bmatrix} \alpha \, e\\ \left(\epsilon-\alpha\mu_N\right)
\end{bmatrix} \, \left(f_N^{\alpha}(\epsilon)-f_j^{\beta}(\epsilon)\right) \, \abs{\mathcal{S}_{Nj}^{\alpha\beta}(\epsilon)}^2.
\end{equation}
Here the factor 2 accounts for spin degeneracy, $e$ is the electrical charge, and we define the zero of energy as being that of the electrochemical potential of the superconductor, i.~e.~$\mu_S=0$. $\mu_N$ is the electrochemical potential of the normal contact~\cite{BTK1982,Benenti2017a}. The labels $\alpha,\beta = \pm$ accounts for the electron-like (+) or hole-like (-) quasiparticle, $f_S^{\alpha}(\epsilon)$ and $f_N^{\alpha}(\epsilon)$ are respectively the energy-dependent Fermi distributions characterizing the superconducting and normal contacts. Depending on whether the quasiparticle behaves as an electron ($\alpha = +$) or a hole ($\alpha = -$), the electrochemical potential $\mu_i$ of lead $i = N,S$ has to be subtracted (electron) or added (hole) to the energy $\epsilon$:
\begin{equation}
f^\alpha_i(\epsilon)=\{e^{(\epsilon-\alpha\mu_i)/k_BT_i}+1\}^{-1},
\end{equation}
with $T_i$ the temperature of lead $i=N,S$ and $k_B$ the Boltzmann constant. 
In Eq.~\eqref{current_Lambert_general}, $\mathcal{S}$ represents the scattering matrix of the whole device, which we now make explicit.

\subsection{Scattering matrix of a N-S AB ring}

The total scattering matrix (s-matrix) $\mathcal{S}$ for this N-S AB ring is the composition of the s-matrix for the AB ring  $S_{AB}$ and the s-matrix of the N-S interface $S_{NS}$ through established rules that conserve the number of quasiparticles flowing through the device \cite{Datta1997, Gresta2021}:
\begin{align}
\label{STOTmain}
\mathcal{S}= S_{AB} \circ S_{NS}.
\end{align}
The s-matrix $S_{NS}$ must account for Andreev reflections \cite{Andreev1964}, allowing an electron to be reflected as a hole and vice-versa. Hence, for a two-terminal device with a single channel, it imposes the form of a $4 \times 4$ matrix to $\mathcal{S}, S_{AB}$ and $S_{NS}$. As the AB ring does not involve any superconducting material, $S_{AB}$ will take a block-diagonal form, indicating that the AB ring does not couple electrons and holes. The matrix $S_{AB}$ depends in general on energy, denoted $\epsilon$ below, and can then be written as:
\begin{equation}
\label{S_AB}
S_{AB}(\epsilon) = \left(
\begin{array}{cccc}
 r^{ee}_{\text{AB}}(\epsilon ) & 0 & t'^{ee}_{\text{AB}}(\epsilon ) & 0 \\
 0 & r^{hh}_{\text{AB}}(\epsilon ) & 0 & t'^{hh}_{\text{AB}}(\epsilon ) \\
 t^{ee}_{\text{AB}}(\epsilon) & 0 & r'^{ee}_{\text{AB}}(\epsilon ) & 0 \\
 0 & t^{hh}_{\text{AB}}(\epsilon )& 0 & r'^{hh}_{\text{AB}}(\epsilon )
\end{array} \right)\,.
\end{equation}
$S_{AB}$ results from the combination of the s-matrices of the central ring and of the two T-junctions, see~\ref{appendix_ring} for all details. For a normal metal, the reflection and transmission amplitudes for electrons and holes in Eq.~\eqref{S_AB} are simply related through electron-hole conjugation rules:
\begin{equation}
\label{particle_hole_transformations}
 r^{hh}_{\text{AB}}(\epsilon ) =  \big(r^{ee}_{\text{AB}}(-\epsilon ) \big)^* ;\quad
 t^{hh}_{\text{AB}}(\epsilon ) =  \big(t^{ee}_{\text{AB}}(-\epsilon ) \big)^*,
\end{equation}
and similarly for the amplitudes $r'$ and $t'$ for incoming waves from the right. In contrast to $S_{AB}$, the s-matrix $S_{NS}$ is characterized by non-vanishing off-diagonal elements to account for Andreev processes (allowing an electron to be reflected/transmitted as a hole and vice-versa):
\begin{equation}
\label{S_NS}
S_{NS} = \left(
\begin{array}{cccc}
 r_{ee} & r_{eh} & t_{e \tilde{e}} & t_{e \tilde{h}} \\
 r_{he} & r_{hh} & t_{h \tilde{e}} & t_{h \tilde{h}} \\
 t_{\tilde{e} e} & t_{\tilde{e} h } & r_{\tilde{e}\tilde{e}} & r_{\tilde{e} \tilde{h}} \\
 t_{\tilde{h} e} & t_{\tilde{h} h} & r_{\tilde{h} \tilde{e}} & r_{\tilde{h}\tilde{h}}
\end{array} \right)\,.
\end{equation}
The reflection and transmission amplitudes for electron, holes and Andreev processes are derived within a Bogoliubov-de Gennes (BdG) formalism to capture the essence of superconductivity \cite{BTK1982,Blasi2020}. The amplitudes $r_{\alpha \beta}$ and $t_{\alpha \beta}$ represent respectively the reflection and transmission amplitudes of an incoming quasiparticle of type $\beta=e,h$ ($\beta=\tilde{e},\tilde{h}$) injected from lead $N$ ($S$) to end up as a quasiparticle of type $\alpha=e,h$ ($\alpha=\tilde{e},\tilde{h}$) in lead $N$ ($S$). For clarity, the tilde denotes the quasiparticles in the superconducting lead, $\tilde{e}$ for electron-like and  $\tilde{h}$ for hole-like quasiparticles, see Fig.~ \ref{System_main}(b). 
Let us remark that in the limit of a vanishing superconducting gap, $\Delta \rightarrow 0$, one recovers the scattering amplitudes derived in previous references investigating a AB ring in a two-terminal device with two normal contacts \cite{Haack2019, Haack2021}. 

Usually, in the literature, the amplitudes in Eq.~\eqref{S_NS} are obtained in the so-called Andreev approximation regime, defined when the superconducting gap is much smaller than the Fermi energy, $\Delta \ll \epsilon_F$~\cite{BTK1982,Pershoguba2019a}. In this case, the scattering amplitudes take a simple analytical form that depends explicitly on $\Delta$ and on the transparency coefficient $Z$, as we detailed in \ref{appendix_NS}.
However, it is important to notice that, within the Andreev approximation, all the details about the curvature of the BdG eigenspectrum are lost and, as consequence, thermoelectric effects get strongly suppressed. 
Therefore, to evidence the thermoelectric feature of our hybrid superconducting system, in the following section we will go beyond the Andreev approximation. Remarkably, an analytical form of the scattering coefficients of Eq.~\eqref{S_NS} can be found beyond the Andreev approximation in the case $Z=0$ -- see~\ref{appendix_C} for details.


\section{Thermoelectric properties of the N-S AB ring}


Thermoelectricity -- i.e. the ability to convert a heat current into an electrical current (Seebeck effect), or vice versa (Peltier effect) -- requires an energy asymmetry between electron-like and hole-like quasiparticles, see Ref.~\cite{Benenti2017a} for a pedagogical review. The Seebeck coefficient $S$ assesses the thermovoltage $\Delta V_{th}$ developed by the device in response to a temperature bias at zero charge current, hence expressed in units of $[V/K]$:
\begin{equation}
\label{Seebeck_general_form}
S=\left.\frac{\Delta V_{th}}{\Delta T}\right|_{I_N=0}\,.
\end{equation}
In a N-S two terminal device within the Andreev approximation, the wave vector amplitudes for electrons and holes are both evaluated at the Fermi energy, $k_{e/h}\simeq k_F$, preventing any possible electron-hole asymmetry in the transmission probability. This results in the absence of any thermoelectric effects. Beyond the Andreev approximation, there exists an electron-hole asymmetry in the wave vectors, which should lead to a thermoelectric response of the N-S junction. To the best of our knowledge, this has not yet been shown in earlier works, and we take the opportunity of this work to fill this gap~\footnote{We report the recent work~\cite{Mukhopadhyay2021} in which authors discuss thermoelectric response beyond Andreev approximation up to the first order in $\epsilon/\epsilon_F$.}. In contrast to the N-S junction, the AB ring hosts quantum interferences that favor thermoelectric effects~\cite{Haack2019,Haack2021,Hofer2014, Samuelsson2017,Kim2003,Blasi2020a}.

We show below that the combination of the AB ring and the N-S junction become beneficial to each other, and allow a thermoelectric response of the order of thousands $\mu V/K$. We investigate the thermoelectric response of the device both in the linear and non-linear response regimes, within and beyond the Andreev approximation. 

%
%

\subsection{Thermoelectric linear response}

The linear response regime is defined when $\Delta T=T_N - T_S \ll T \equiv (T_N + T_S)/2$ for the temperature bias and $e\Delta V=(\mu_N - \mu_S) \ll k_B T$ for the voltage bias~\cite{Benenti2017a}. In this situation, the electrical charge current ($I_N$) and the heat current ($J_N$) in Eq.~\eqref{current_Lambert_general} are simply expressed in terms of the Onsager matrix $\vb{L}$ \cite{Benenti2017a}:
\begin{equation}
\label{Onsager_PRL}
\left(
\begin{array}{c}
I_N \\
J_N 
\end{array} \right) 
= 
\left( \begin{array}{cc}
L_{11} & L_{12} \\
L_{21} & L_{22}
\end{array} \right) 
\left( \begin{array}{c}
\Delta V/T \\
\Delta T/T^2\end{array} \right) \,.
\end{equation}
The diagonal Onsager coefficients determine the electrical ($L_{11}$) and heat ($L_{22}$) conductance, whereas the off-diagonal terms $L_{12}=L_{21}$ set the thermoelectric properties of the device in the linear response regime.\\

\textit{Seebeck coefficient --} 
In the linear response regime, the Seebeck coefficient defined in Eq.~\eqref{Seebeck_general_form} is determined by the Onsager coefficients~\cite{Benenti2017a}:
\begin{equation}
S=\frac{1}{T}\frac{L_{12}}{L_{11}}.
\end{equation}
\begin{figure}[h]
	\centering
	\includegraphics[width=1\textwidth]{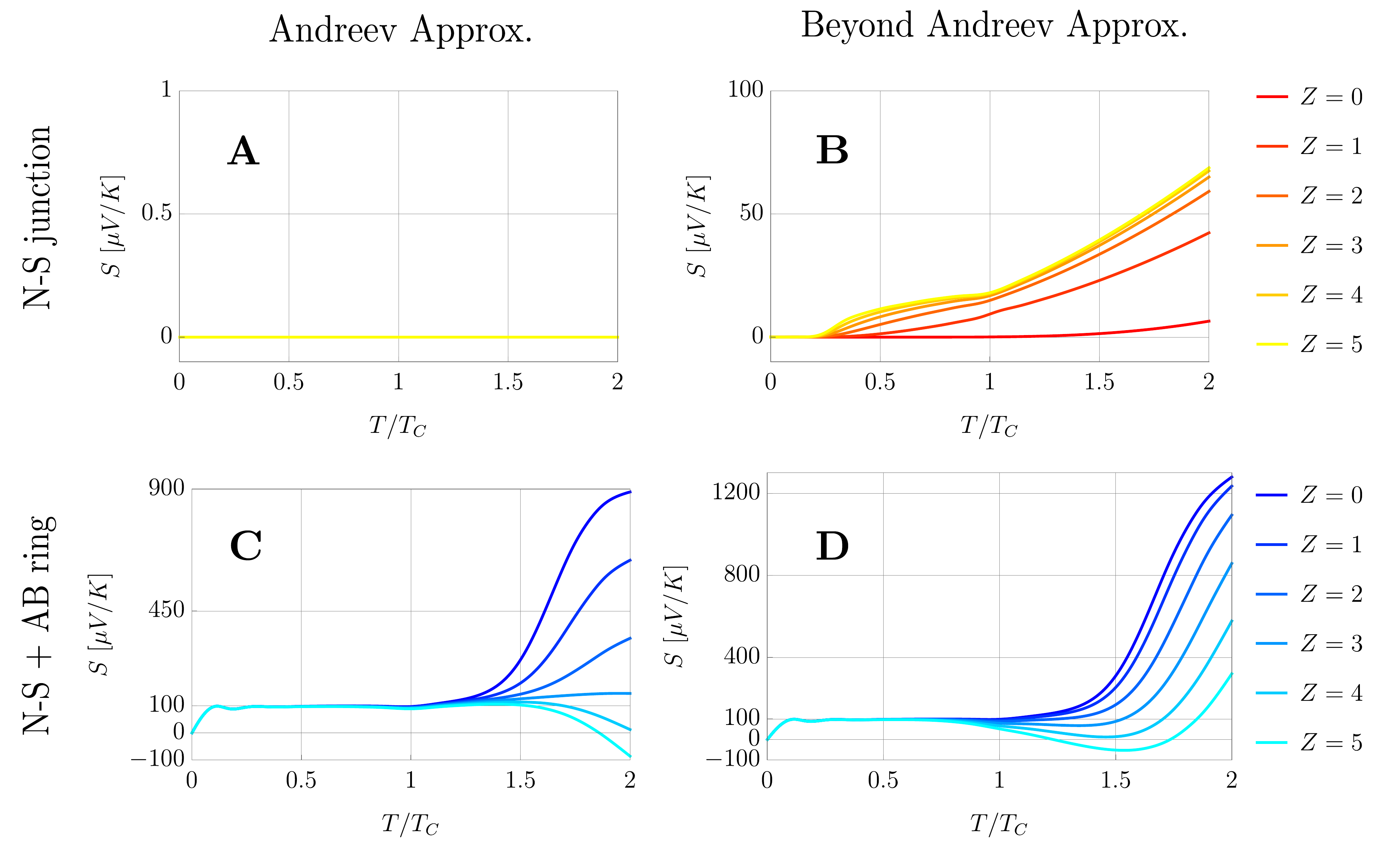}
 	\caption{Seebeck coefficient (in units of $\mu V/K$) as a function of the mean temperature $T$ (in units of the critical temperature $T_C$) for different values of the barrier strength $Z$. Upper panels show the case of the N-S junction \textit{without} the AB ring, bottom panels show the configuration of the N-S junction \textit{with} the AB ring. Left panels have been obtained in the Andreev approximation ($\epsilon,\Delta\ll \epsilon_F$), right panels have been obtained beyond the Andreev approximation limit with $\epsilon_F=5\Delta$. Here we considered $k_F\xi=1$ (where $\xi=\hbar v_F/\Delta$ is the superconducting coherence length), $\tau=0.5$, while other parameters have been optimized in the temperature range $T/T_C\in [0,2]$ and are: $L_u/\xi=L_d/\xi=0.1$, $V_g/\pi\Delta=0$ and $2\pi\Phi/\Phi_0=0.5$.}
	\label{AAvsBAA}
\end{figure}
Figure~\ref{AAvsBAA} shows the Seebeck coefficient of the N-S AB ring in the linear response regime within and beyond the Andreev approximation, and compares it to the Seebeck coefficient for the N-S junction alone, without the AB ring. The Seebeck coefficient is plotted as a function of the temperature $T/T_C$ for different values of $Z$. Other parameters have been optimized using the method of the  gradient descent in order to get the highest value of the Seebeck coefficient in the temperature range $T/T_C\in[0, 2]$. To account for the temperature dependence of the superconducting order parameter, we use the following relation for the energy gap:
\begin{equation}
\label{Gap}
\Delta(T)=\Delta(0)\tanh{\left(1.74\sqrt{\frac{T_C}{T}-1}\right)}
\end{equation}
which is accurate better than $2\%$ with respect to the self-consistent BCS result \cite{Tinkham1966,Kamp2019}. Panel \textbf{A} evidences the absence of any thermoelectric response of the N-S junction within the Andreev approximation with $S=0$ for all values of $Z$  (as explained earlier, in this limit no electron-hole asymmetry exists). Beyond the Andreev approximation, the situation clearly differs, see panel \textbf{B}. The Seebeck coefficient is finite and increases monotonically both as a function of temperature and $Z$ reaching values of the order of $\sim 60 \mu V/K$. Interestingly, it decreases when going to the limit of a transparent barrier, $Z \rightarrow 0$. To understand this dependence, we investigated numerically the case $Z=0$. In this limit, the scattering amplitudes in $S_{NS}$ obtained beyond the Andreev approximation weakly deviate from that given in the Andreev approximation only for $\epsilon\gtrsim\Delta$ within few $k_BT$. As a consequence, at low temperatures the thermoelectric response is weak, and increases by increasing temperature (see red curve in panel \textbf{B}).

In panels \textbf{C} and \textbf{D}, we first observe a plateau, the Seebeck coefficient remains constant at $\sim 100 \mu V/K$, when the gap is open, i.e. for $T < T_C$. Interestingly, this plateau is independent of the value of Z. 
At $T \approx T_C$ (when the gap is closed), the Seebeck coefficient starts to deviate until it reaches (at $T=2 T_C$) values of the order of $\sim 900~\mu V/K$ in the Andreev approximation (panel \textbf{C}) and $\sim 1300~\mu V/K$ beyond the Andreev approximation (panel \textbf{D}).
Let us remark that in the regime beyond the Andreev approximation, we verified numerically that the Seebeck coefficient increases linearly with temperature for the N-S junction, while it saturates to a constant value $\sim 1400 \, \mu V/K$ for the N-S AB ring. We also note that $S$ is maximal at $Z=0$ (ideal contact with the superconductor) in presence of the AB ring (panels \textbf{C} and \textbf{D}), whereas - as already discussed - it decreases with $Z$ in presence of the N-S junction only (beyond Andreev approximation, panel \textbf{B}).\\


\begin{figure}[t]
\centering
	\includegraphics[width=0.8\textwidth]{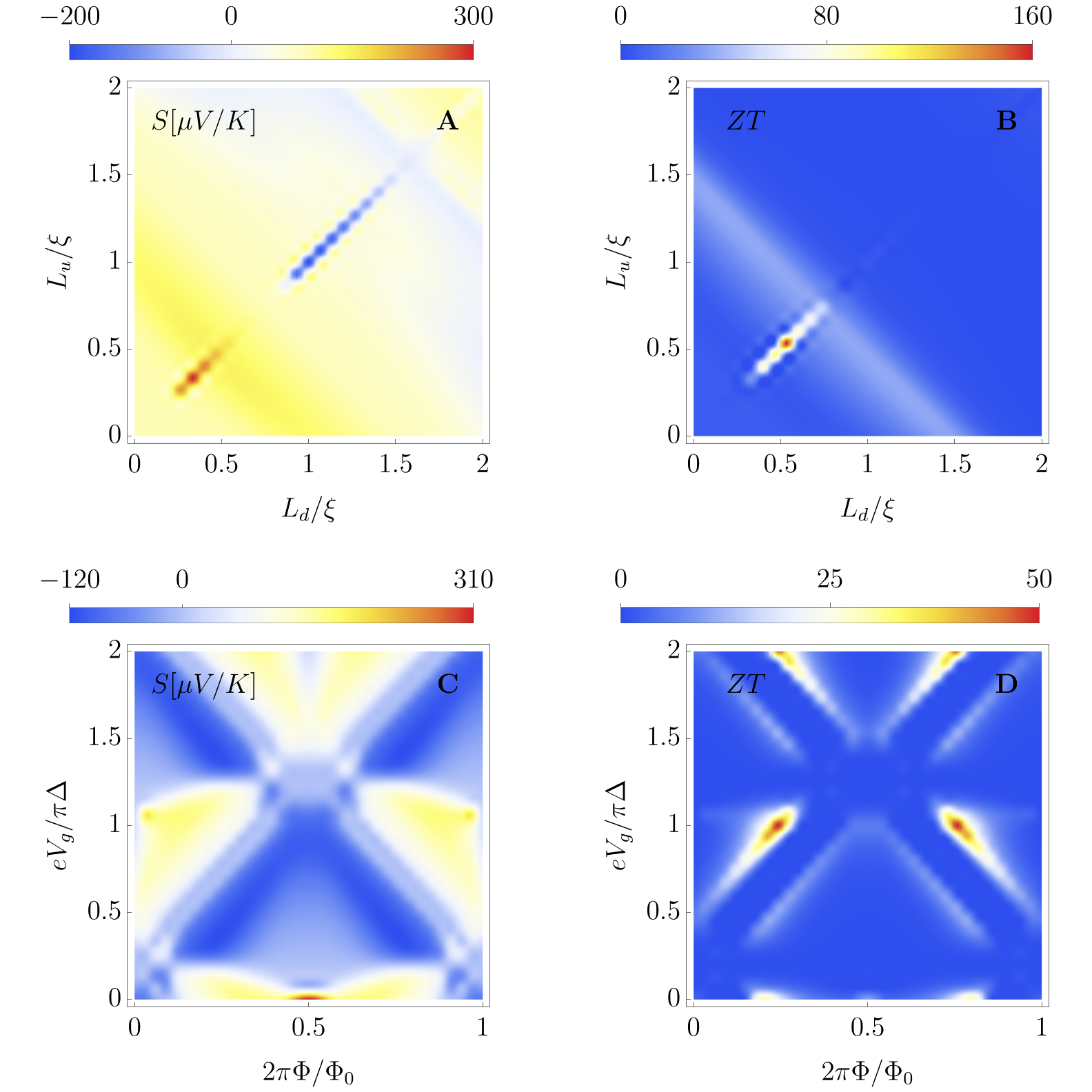}
 	\caption{Seebeck coefficient (left panels) in units of $\mu V/K$ and the figure of merit $ZT$ (right panels) as function of the upper and downer branches lengths $L_u/\xi$ and $L_d/\xi$ of the AB ring expressed in units of the coherence length $\xi=\hbar v_F/\Delta$ (upper panels) and as a functions of magnetic flux $2\pi\Phi/\Phi_0$ and the gate potential $V_g/\pi\Delta$ (bottom panels). For upper panels we fixed $V_g/\pi\Delta=0$ and $2\pi\Phi/\Phi_0=0.5$, for bottom panels we considered $L_u/\xi=L_d/\xi=0.3$: these choices maximize the value of the Seebeck coefficients. All panels plots have been obtained in the Andreev approximation where we considered $\tau=0.1$, $Z=0.1$ and $k_F\xi=1$.} 
	\label{SZT}
\end{figure}

\textit{Tunable thermoelectric properties --} We now characterize the thermoelectric response and the efficiency of this N-S AB ring as a function of key parameters for an experiment: external gate voltage $V_g$ applied onto one of the arms of the AB ring, length imbalance $\Delta L=L_u-L_d$ between the two arms and the AB flux $\Phi$ due to an external magnetic field applied perpendicularly to the sample. Panels \textbf{A} and \textbf{C} of Fig.~\ref{SZT} show density plots of the Seebeck coefficient as a function of these parameters. In the N-S AB ring device, it seems advantageous for increasing $S$ to operate with a balanced AB ring, with symmetric arms $L_u=L_d$. In this situation, the behaviour of $S$ as a function of external gate voltage $V_g$ and AB flux $\Phi$ highlights some optimal values to reach high value for $S$, about 300 $\mu V/K$. Here we have considered a mean temperature $T/T_C=0.5$ such that the gap of the superconducting lead is open, and a transmission probability for the T-junctions $\tau=0.1$. The other parameters have been optimized by using the method of the gradient descent in order to get the higher value of the Seebeck coefficient. The right column (panels \textbf{B} and \textbf{D}) shows the ZT coefficient, a figure of merit for assessing the maximal efficiency $\eta_{max}$ of a thermoelectric device in the linear response regime. It approaches the Carnot efficiency $\eta_C=\Delta T/T$ for $ZT\rightarrow\infty$ \cite{Benenti2017a}. In terms of the Onsager matrix and coefficients introduced in Eq.~\eqref{Onsager_PRL}, it reads:
\begin{equation}
\eta_{max}=\eta_C\frac{\sqrt{ZT+1}-1}{\sqrt{ZT+1}+1} \quad \text{with} \quad ZT=\frac{L_{12}^2}{\text{Det}[\vb{L}]}\,.
\end{equation}

As it emerges from the top panels of Fig.~\ref{SZT}, both the Seebeck coefficient and the $ZT$ figure of merit are sizable only when the lengths of the upper and lower branches of the AB ring are equal $L_u=L_d$, corresponding to the darker stripe on the diagonal of the density plots. More precisely, the Seebeck coefficient takes its maximal value when $L_u=L_d\approx 0.3$, in which case $S\approx 300 \mu V/k$ and $ZT\approx 100$, the latter being one order of magnitude bigger than the one found in Ref.~\cite{Haack2019}.
This excellent thermoelectric response in the linear regime motivates the investigation of its thermoelectric properties in the non-linear regime. As direct application, we exploit them to propose the N-S AB ring as efficient quantum heat engine in the following section.

\subsection{Quantum heat engine in the non-linear response regime}

\begin{figure}[!htb]
	\centering
	\includegraphics[width=1\linewidth]{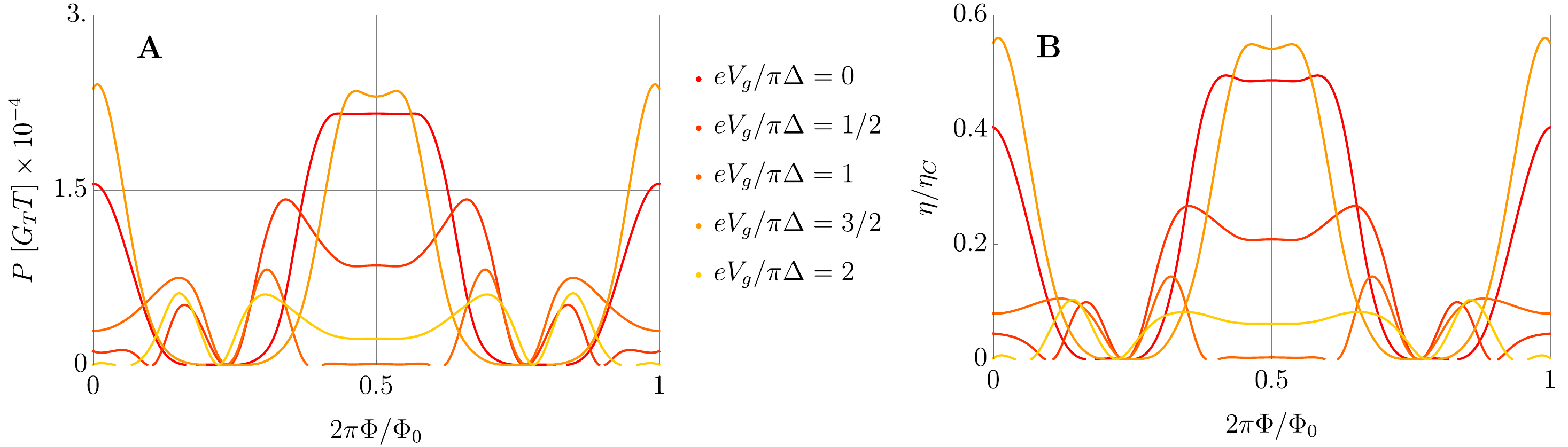}
 	\caption{Power in units of $G_TT$, with $G_T=(\pi^2/3h)k_B^2 T$ the thermal conductance quantum (panel \textbf{A}), and the efficiency $\eta$ in units of the Carnot efficiency $\eta_C=\Delta T/T$ (panel \textbf{B}) as function of the magnetic flux $\Phi$ in units of the flux quantum $\Phi_0$ for different values of the gate voltage $eV_g/\pi\Delta$. Here the other parameters used to maximize the efficiency are: $R_L/(h/2e^2)=100$, $Z=0.04$, $\tau=0.24$, $L_u/\xi=0.1$, $L_d/\xi=1.26$, $T/T_C=0.5$ and $\Delta T/T_C=0.1$.} 
	\label{P_eta}
\end{figure}

Predicting a high Seebeck coefficient for a given nanoscale device is extremely relevant for designing efficient quantum thermal engines. Indeed, by closing the circuit with a load resistance $R_L$ (see Fig.~\ref{System_main} (a)) connected with superconducting wires to the device (no Joule dissipation in the wires), one can generate an electrical power $P$ from the thermovoltage developed by the device in response to the thermal bias, $P=\frac{\Delta V_{th}^2}{R_L}$. The efficiency of such a device is then the ratio of the electrical power (output) over the heat current from the hot contact in the stationary state evaluated at the thermovoltage (input), $\eta=\frac{P}{J_N(\Delta V_{th})}$.

In the non-linear regime, the thermovoltage developed by the device has to be calculated by solving the following equation for the closed circuit (we refer to \cite{Haack2021} and references therein for more details):
\begin{equation}
I_N\left(\Delta V_{th}\right)=-\frac{\Delta V_{th}}{R_L}.
\end{equation}
We evaluate numerically the thermovoltage in the closed circuit configuration as a function of the magnetic flux $\Phi$ and of the gate voltage $V_g$, showing the results in Fig.~\ref{P_eta}. Panel \textbf{A} corresponds to the generated power $P$ and panel \textbf{B} to the efficiency $\eta$.
We predict values for the efficiency at maximum power up to $\sim 55\%$ of the Carnot efficiency. 
Compared to the values $\eta \sim 40\%$ obtained with a normal AB structure~\cite{Haack2021}, we can associate the higher efficiency to the N-S interface as it leads to a decrease of the heat current entering the denominator of $\eta$. Indeed, the N-S junction acts as a mirror for the heat current: Andreev reflections block energy and heat fluxes  since the latter are propagated by single quasi-particles and not by Cooper pairs in the condensate~\cite{Benenti2017a}.
We note that power and efficiency present the same behavior as a function of $2\pi\Phi/\Phi_0$ and $V_g$, which implies that maximum efficiency and maximum generated power are reached for the same values of parameters. 

\section{Hybrid AB ring as quantum thermal rectifier}

Demonstrating thermal rectification with high efficiency at the quantum scale is currently a very active research direction, both from a fundamental point of view and in applied physics for quantum engineering. In a two-terminal setup subject to thermal bias, thermal rectification is achieved when left-to-right currents differ upon exchanging the temperatures of the two contacts. 
From the theory of superconductivity, it is clear that an efficient way to break left-right symmetry is to exploit the dependence of the density of states (DOS) of superconductors as a function of temperature. Indeed, as recalled in Eq.~\eqref{Gap}, the superconducting gap, and consequently the DOS, changes with temperature~\cite{Casati2007,Roberts2011,Martinez-Perez2013,Giazotto2013}. Upon exchanging the temperatures, DOS of the contacts are modified, inducing different heat currents in the two configurations. Hence, thermal rectification at the nanoscale was predicted and measured in N-S junctions in the past decade \cite{fornieri2017towards}. 

To investigate and characterize this operating mode for the hybrid N-S AB ring device, we consider again the open-circuit setup (i.e.~without the load resistance) beyond the linear response regime. 
Specifically, we consider two configurations for the thermal bias. In the \textit{forward} configuration, a thermal gradient is created by setting $T_N=T_{hot} > T_S = T_{cold}$, leading to a total heat current $J_N^+$ flowing from $N$ to $S$. In the \textit{reverse} thermal bias configuration, the thermal gradient is inverted, $T_N=T_{cold} < T_S = T_{hot}$, leading to a heat current $J_N^-$ flowing from $S$ to $N$. 
It follows that thermal rectification is achieved whenever $\abs{J_N^+} \neq \abs{J_N^-}$. Let us remark that the thermovoltage developed by the N-S AB ring in response to a temperature gradient \textit{must} be taken into account when calculating the heat currents $J_N^+$ and $J_N^-$.
The thermovoltages in the two configurations are solutions of the following equations for the charge current:
\begin{equation}
\label{Vth_equation}
\left\{\begin{array}{l}
I_{N}\left(\Delta V_{th}^+,T_N=T_{hot},T_S=T_{cold}\right)=0 \\
I_{N}\left(\Delta V_{th}^-,T_N=T_{cold},T_S=T_{hot}\right)=0
\end{array}\right.
\end{equation}
with the charge current given by Eq.~\eqref{current_Lambert_general}.\\

\begin{figure}[t]
	\centering
	\includegraphics[width=1\linewidth]{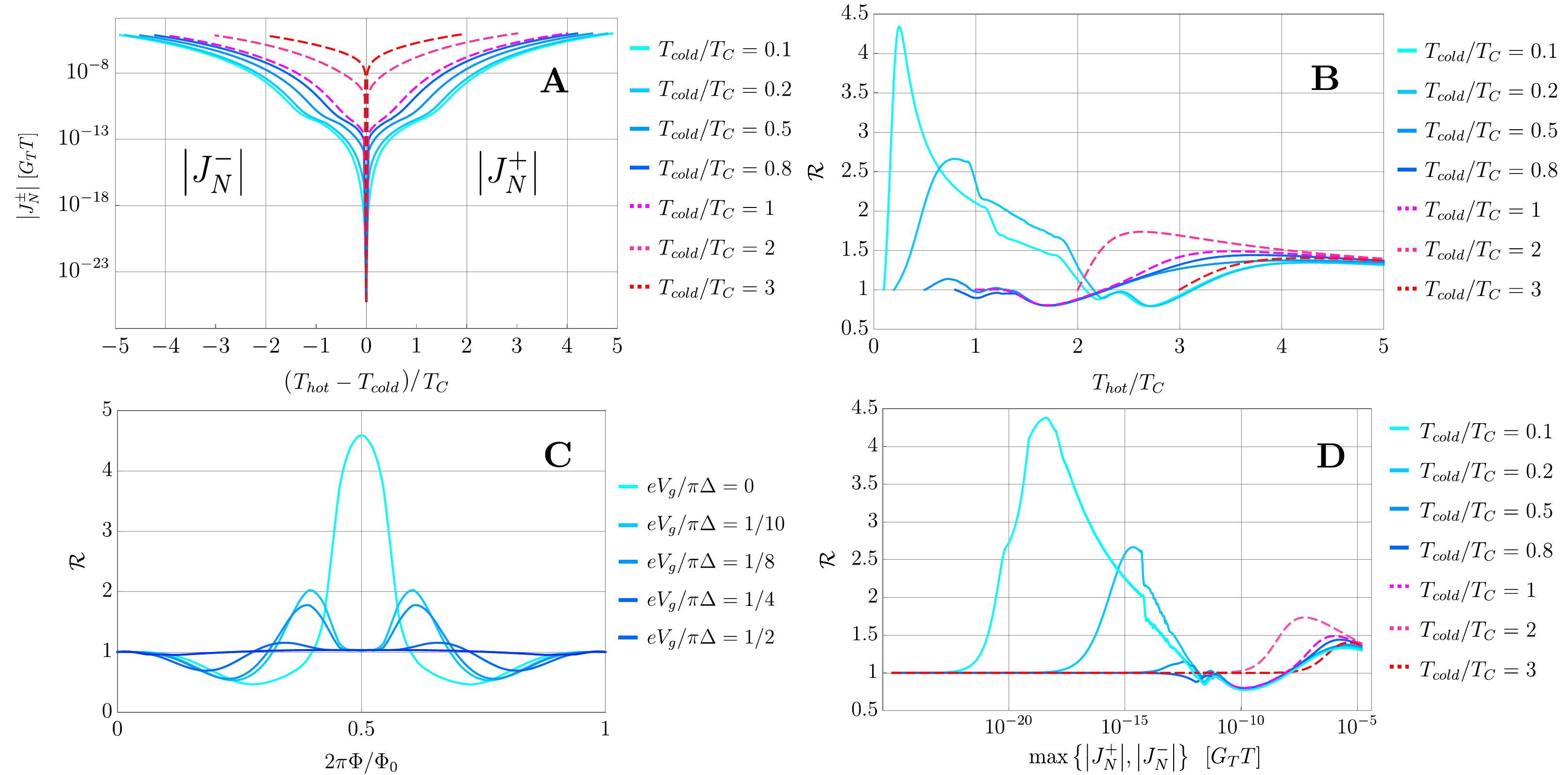}
 	\caption{Rectification with the hybrid AB ring. Panel \textbf{A}: Log-scale plot of the absolute value of the heat current $|J_N^\pm|$ in units of $G_TT$ as a function of $(T_{hot}-T_{cold})/T_C$. Right branch corresponds to $\abs{J_N^+}$ in the forward configuration while left branch represents $\abs{J_N^-}$ in the reverse configuration. Blueish solid curves represent the case in which $T_{cold}/T_C<1$ (the gap is open in the $S$ contact), reddish dashed curves have been obtained for $T_{cold}/T_C\geq1$ (both contacts behave as normal metals). Panel \textbf{B}: Rectification factor $\mathcal{R}$ as function of $T_{hot}/T_C$ for different values of $T_{cold}/T_C$ with $V_g = 0$ and $2\pi\Phi/\Phi_0=0.5$ . Panel \textbf{C} : Rectification factor $\mathcal{R}$ as function of $2\pi\Phi/\Phi_0$ for different values of $eV_g/\pi\Delta$ at fixed values of $T_{cold}/T_C=0.1$ and $T_{hot}/T_C=0.15$. Panel \textbf{D} : parametric plot of $\mathcal{R}$ and $\max{\left\{\abs{J_N^+}, \abs{J_N^-}\right\}}$ as a function of $T_{hot}/T_C$ for different values of $T_{cold}/T_C$. Fixed parameters: $\tau=0.1$, $Z=0.2$, $k_F\xi=1$, $L_u=L_d=0.1$. } 
	\label{Rectification}
\end{figure}

Panel \textbf{A} in Fig.~\ref{Rectification} shows in log-scale the absolute value of the heat currents $|J_N^\pm|$ as a function of $(T_{hot}-T_{cold})/T_C$ for different values of $T_{cold}$. 
Note that for $T_{cold}/T_C<1$ (see blueish solid curves), we probe the superconducting properties of the $S$ contact, while for $T_{cold}/T_C\geq1$ (see reddish dashed curves) both contacts behave as normal metals.
Rectification can be assessed by its figure of merit, the rectification factor defined as:
\begin{equation}
\mathcal{R}=\abs{J_N^+}/\abs{J_N^-}.
\end{equation}
If $\mathcal{R}>1$, the heat current flows from left to right ($N\rightarrow S$), while if $\mathcal{R}<1$ the heat current flows from right to left ($N\leftarrow S$). If $\mathcal{R}=1$, the heat current does not have a specific direction, i.~e.~there is no rectification.
Panel \textbf{B} in Fig.~\ref{Rectification} represents the rectification factor $\mathcal{R}$ as function of $T_{hot}/T_C$ for different values of $T_{cold}/T_C$.
Here we considered the hot temperature in the range $T_{cold}\leq T_{hot}\leq 5T_C$, while for larger temperatures, i.e.~$T_{hot}\gg T_C$, we verified that $\mathcal{R}$ approaches 1. As can be noticed in panel \textbf{B}, the higher rectification occurs for small values of $T_{cold}\ll T_C$, while it decreases by increasing $T_{cold}$.
In particular, a value for rectification of about $\mathcal{R}\simeq 4.5$ is achieved for $T_{cold}/T_C=0.1$ and $T_{hot}/T_C=0.15$: this corresponds to a value of the forward heat current ($J_N^+$) greater than $350\%$ compared to the reverse one ($J_N^-$). 
Furthermore, it is important to notice that, differently from systems that do not exhibit thermoelectric properties, in our case rectification occurs even for $T_{cold}> T_C$, namely when the gap of the superconductor is closed and the system behaves as an effective $N$-$AB$-$N$ device.
This happens because, although in such an effective $N$-$AB$-$N$ regime the system is left-right symmetric (namely the scattering matrix $\mathcal{S}$ is not affected by the exchange of temperatures, $T_{cold}\leftrightarrow T_{hot}$), the thermovoltage generated in the forward configuration is different from that obtained in the reverse one, i.e.~$\Delta V_{th}^+\neq \Delta V_{th}^-$, as results by explicitly solving Eqs. \eqref{Vth_equation}. This results in a forward heat current different from the reverse one, $\abs{J_{N}^+(\Delta V_{th}^+)}\neq \abs{J_{N}^-(\Delta V_{th}^-)}$, which reflects into a finite rectification. However, it is important to stress that the rectification obtained when the system behaves like an effective $N$-$AB$-$N$ device (i.e. when $T_{cold}>T_C$) is at least 3 times smaller than the one obtained when the gap is open and the right lead is still superconducting (i.e. when $T_{cold}<T_C$), as can be seen by comparing the cyan solid curve with the violet dashed one in panel \textbf{B} of Fig.~\ref{Rectification}.

We also investigate the tunability of this rectifier by means of the external parameters provided by the N-S AB ring device. 
Panel \textbf{C} in Fig.~\ref{Rectification} shows $\mathcal{R}$ as a function of the magnetic flux $\Phi$ for different values of the gate voltage $V_g$. The temperatures $T_{cold}/T_C=0.1$ and $T_{hot}/T_C=0.15$ have been fixed to optimize $\mathcal{R}$. It becomes clear that this device is fully phase-tunable. Controlling the magnetic flux $\Phi$ allows us to turn on and off the rectification ability of the N-S AB ring. \\

Finally, we would like to emphasize that a quantum heat rectifier should also be characterized by good heat conduction properties to constitute a useful device. Indeed, as recently discussed in \cite{Khandelwal2022}, there typically exits a trade-off between heat rectification and heat conduction. Large rectification factors often occur at low maximum heat current through the device. This motivates us to investigate the heat rectification factor and the heat currents in a complementary way. To this way, we show in panel \textbf{D} a parametric plot of $\mathcal{R}$ and $\max{\left\{\abs{J_N^+}, \abs{J_N^-}\right\}}$ as a function of $T_{hot}/T_C$. While the trade-off between heat rectification and heat conduction is clearly visible (highest $\mathcal{R}$ occurs at lowest $\max{\left\{\abs{J_N^+}, \abs{J_N^-}\right\}}$, see cyan solid curve corresponding to $T_{cold}/T_C=0.1$), the numerical results predict sizable rectification of $50\%$ happening at several tens of $[nW]$ for all values of $T_{cold}$.

\section{Experimental implementation of the hybrid AB ring}

Let us now discuss a possible experimental setup to implement the hybrid AB interferometer. Since the structure also contains a superconducting lead, III-V semiconducting alloys like InAs \cite{giazotto2004josephson,amado2014ballistic,amado2013electrostatic,fornieri2013ballistic} or In$_{0.80}$Ga$_{0.20}$As \cite{deon2011proximity,deon2011quantum} two-dimensional electron gases (2DEGs) are suitable candidates for the realization of the ballistic loop structure.  
These materials typically provide Schottky barrier-free contacts with metals, which is a crucial requirement in order to achieve highly-transparent N-S interfaces, and thereby maximize Andreev reflection at the contact with the superconductor. 
Such III-V 2DEGs can be easily gated by means of side or top gates in order to finely tailor the details of the thermoelectric AB structure. As far as the superconducting element is concerned, aluminum (Al, providing an energy gap of $\sim 200\mu$eV and critical temperature around 1.4K) or niobium (Nb, providing an energy gap of $\sim 1.5$meV  and critical temperature $\sim 9$K) thin layers are ideal superconductors to be coupled to the 2DEG AB ring. 

The AB quantum thermal structure analyzed so far needs to be thermally-biased in order to provide either thermoelectric response or heat rectification properties \cite{fornieri2017towards}. To this end superconducting tunnel junctions (typically made of oxidized Al layers) can be integrated in the normal and superconducting leads forming the structure \cite{giazotto2006opportunities}, and can be used as electron heaters  so to impose a suitable thermal gradient across the structure via Joule heating, or can be used to measure the quasiparticle temperature thereby operating as sensitive electron thermometers \cite{giazotto2006opportunities}. On the one hand, the thermoelectric response of the hybrid AB interferometer can be proved by setting a thermal gradient across the ring (from a few tens to a few hundreds mK depending on the average temperature of the structure), and by measuring either the thermovoltage in an open-circuit configuration or the thermocurrent by closing the circuit upon a suitable loading resistor \cite{fornieri2017towards}. 
On the other hand, the heat rectification character of the system can be demonstrated by tunnel-coupling two identical normal metal (N) reservoirs to the N-S AB structure \cite{fornieri2015electronic,martinez2015rectification}, one on the left and the other on the right, each of them equipped with superconducting tunnel junctions thereby implementing electron heaters and thermometers. The forward thermal bias configuration can be achieved by intentionally increasing the electronic temperature up to $T_{hot}$ in one of the two N electrodes (i.e., up to several hundreds mK to achieve the full non-linear regime in temperature), and by measuring the resulting steady-state temperature in the opposite electrode. The reverse thermal bias configuration is obtained similarly by simply inverting the heating in the other N electrode. The difference of the two measured temperatures for any given $T_{hot}$  can be used to assess the degree of thermal rectification as a function of electrostatic gating and magnetic flux \cite{fornieri2017towards,martinez2015rectification}.


\section{Conclusions}

In summary, we have theoretically analyzed both the thermoelectric and rectification response of a ballistic interferometer consisting of an Aharonov-Bohm quantum ring coupled to a normal metal and a superconducting lead. As a thermoelectric quantum device, the N-S AB interferometer is able to provide a sizable Seebeck coefficient as large as $100\mu$V/K below the superconducting critical temperature $T_C$, independently of the transmissivity of the ring/superconductor interface. In addition, at  temperatures larger than $T_C$, the Seeebeck coefficient obtains values exceeding 1mV/K for an ideal contact with the superconductor, whereas it is strongly suppressed by decreasing the ring/superconductor interface transmissivity. Yet, in terms of the $ZT$ coefficient, the N-S AB device obtains values as large as $\sim 160$, which appear promising in light of the implementation of efficient quantum heat engines. In such a context, sizable values as large as $\sim 55\%$ of the Carnot efficiency can, in principle, be achieved in the structure under suitable tuning of the gate voltage and the magnetic flux piercing the interferometer. All this confirms the potential of the N-S AB interferometer as a prototypical platform for the realization of efficient quantum thermal machines.

Moreover, the presence of a superconducting lead breaks left-right thermal symmetry in the structure, thereby allowing  finite heat rectification to occur in the interferometer. In particular, rectification coefficients as large as $\sim350\%$ can be achieved upon proper tuning of the structure parameters. 

Finally, as far as the realization of the structure is concerned, III-V semiconducting alloys  such as InAs or In$_{0.80}$Ga$_{0.20}$As realizing two-dimensional electron gases combined to an Al or Nb superconducting electrode are suitable candidates for the realization of the ballistic Aharonov-Bohm hybrid interferometer. The flexibility offered by the above material systems seems indeed ideal in light of the realization of quantum thermoelectric machines implemented through ballistic AB interferometers to be exploited in future quantum technology applications  operating at subkelvin temperatures.

\section*{Acknowledgements}
FG acknowledges the EU’s Horizon 2020 research and innovation program under Grant Agreement No. 800923 (SUPERTED) and No. 964398 (SUPERGATE) for partial financial support. GB and GH acknowledge support from the Swiss National Science Foundation through the NCCR QSIT (GB and GH) and from the NCCR SwissMAP and a starting grant PRIMA PR00P2\_179748 (GH).

\appendix

\section{Scattering matrix of the AB ring}
\label{appendix_ring}

\begin{figure}[!htb]
	\centering
	\includegraphics[width=0.9\columnwidth]{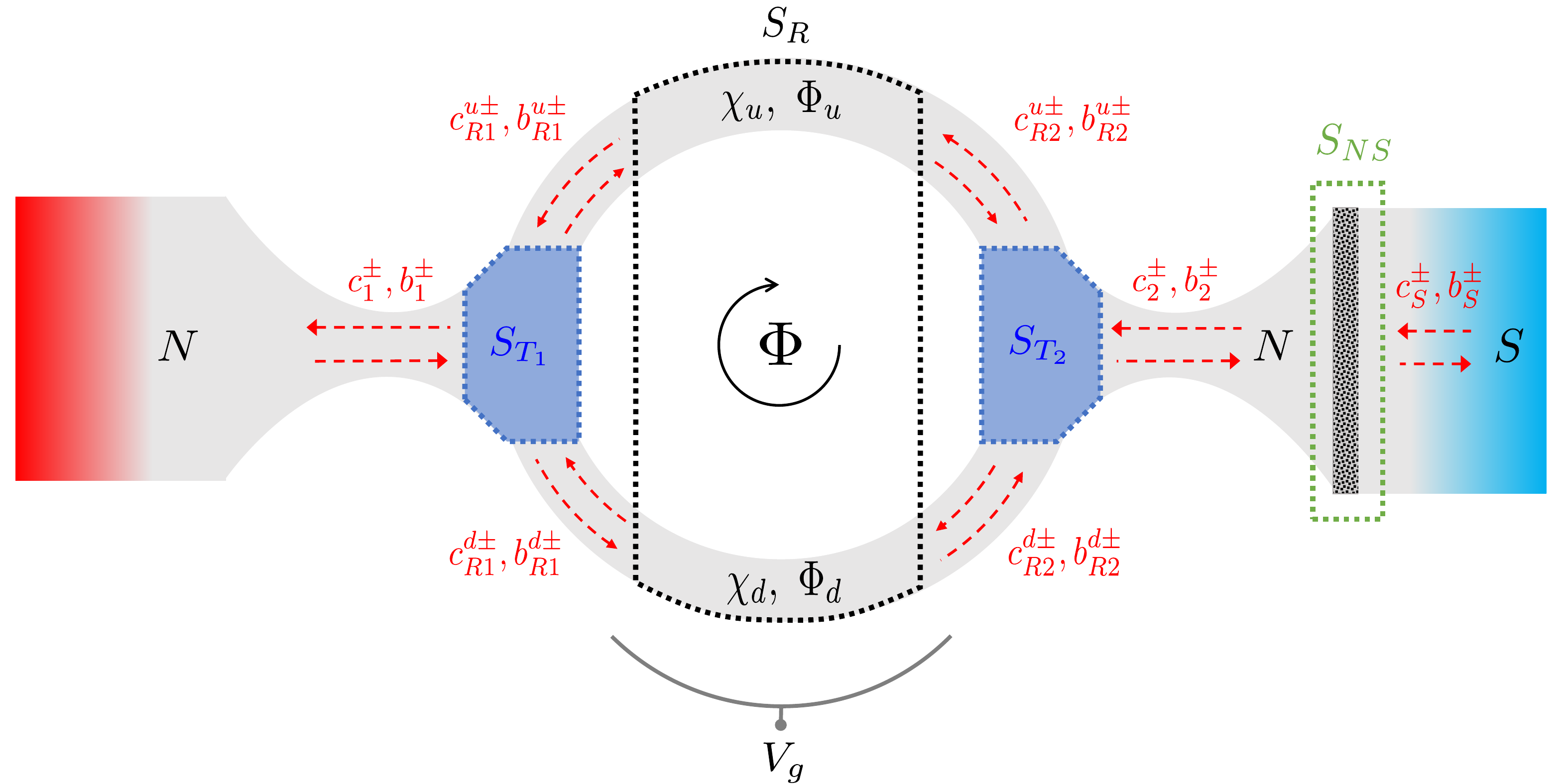}
 	\caption{Sketch of the hybrid AB ring. Dotted lines with different colors encircle respectively the scattering regions of the T-junctions (blue dotted lines), the ring (black dotted line) and the N-S junction (green dotted line). Red dashed arrows indicate the direction of propagation of the incoming and outgoing particles (holes).}
	\label{System}
\end{figure}
Here we provide details on the derivation of the  scattering matrix $S_{AB}$ of Eq.~\eqref{S_AB}.
Specifically, in \ref{sub_app_ST1_ST2_SR}, we proceed by writing first the particle-scattering matrices $S_{T_1,T_2}$ for the T-junctions and $S_R$ for the ring. Then, in \ref{sub_app_comb_scattering_matrix}, we combine them to get the $2\times 2$ particle-scattering matrix of the AB ring connected with the T-junctions, that only describes the scattering processes of particles. Finally, in order to account for the presence of holes, in \ref{sub_app_presence_superconductor_A3} we use the BdG formalism to extend the matrix to the full particle-hole space.

\subsection{T-junctions and AB ring particle-scattering matrices}
\label{sub_app_ST1_ST2_SR}

As shown in Fig.~\ref{System}, we model our setup as a two-terminal geometry consisting of a AB ring connected to two reservoirs through T-junctions \cite{Buttiker1984}. This system can be described via three scattering matrices: $S_{T_1}$ and $S_{T_2}$ describing respectively the left and right T-junctions connecting the AB ring to the contacts, and $S_R$ which describes the ring.
Following Refs.~\cite{Buttiker1984,Haack2019,Haack2021}, the real particle-scattering matrices of the left and right T-junctions are given by
{\small
\begin{equation}
\label{S1}
\left(
\begin{array}{c}
 c^-_1 \\
 c^{u+}_{\text{R1}} \\
 c^{d+}_{\text{R1}} \\
\end{array}
\right)
=
\left(
\begin{array}{ccc}
 -(a_1+b_1) & \sqrt{\tau_1/2} & \sqrt{\tau_1/2} \\
 \sqrt{\tau_1/2} & a_1 & b_1 \\
 \sqrt{\tau_1/2} & b_1 & a_1 \\
\end{array}
\right)_{S_{T_1}}
\left(
\begin{array}{c}
 c^+_1 \\
 c^{u-}_{\text{R1}} \\
 c^{d-}_{\text{R1}} \\
\end{array}
\right)
\end{equation}

\begin{equation}
\label{S2}
\left(
\begin{array}{c}
 c^{u-}_{\text{R2}} \\
 c^{d-}_{\text{R2}} \\
 c^{+}_{2} \\
\end{array}
\right)
=
\left(
\begin{array}{ccc}
 a_2 & b_2 & \sqrt{\tau_2/2} \\
 b_2 & a_2 & \sqrt{\tau_2/2} \\
 \sqrt{\tau_2/2} & \sqrt{\tau_2/2} & -(a_2+b_2) \\
\end{array}
\right)_{S_{T_2}}
\left(
\begin{array}{c}
 c^{u+}_{\text{R2}} \\
 c^{d+}_{\text{R2}}\\
 c^{-}_{2} \\
\end{array}
\right)
\end{equation}
}
where we indicated with $c_{i}^{\pm}$ the incoming and outgoing particles in the left ($i=1$) and right ($i=2$) lead respectively, and with $c_{Rj}^{k\pm}$ the incoming and outgoing particles on the left ($j=R1$) and right ($j=R2$) side of the ring. The index $k=u,d$ labels, respectively, the upper and downer branch of the ring, and $\pm$ indicate the direction of propagation of particles ($+$ for right movers and $-$ for left movers).
In Eqs. \ref{S1} and \ref{S2}, we defined the scattering amplitudes
\begin{align}
a_i&=\frac{1}{2}\left(\sqrt{1-\tau_i}- 1\right)\nonumber\\
b_i&=\frac{1}{2}\left(\sqrt{1-\tau_i}+ 1\right)
\end{align}
where $\tau_i\in[0,1]$ represents the transmission probability of the left ($i=1$) and right ($i=2$) T-junction respectively. For simplicity, in the main text we considered $\tau=\tau_1=\tau_2$ (symmetric T-junctions).
As a further remark, it is useful to notice that, with the choice of basis we did in Eqs. \ref{S1} and \ref{S2}, the scattering matrices of the T-junctions can be written in the standard form
\begin{equation}
 S_{T_i}= 
 \begin{pmatrix}
r_i & t_i^{\prime} \\ 
t_i &  r_i^{\prime}
\end{pmatrix},
\end{equation}
where $r_i$ and $r_i^{\prime}$ are square block matrices concerning reflected particles, whereas $t_i$ and $t_i^{\prime}$ are rectangular block matrices concerning particles transmitted through the left ($i=1$) and right ($i=2$) T-junction respectively.
In the same way we can write the particle-scattering matrix $S_R$ describing the ring, which is given by

\begin{equation}
\label{SR}
\left(
\begin{array}{c}
 c^{u-}_{\text{R1}} \\
 c^{d-}_{\text{R1}}\\
 c^{u+}_{\text{R2}} \\
 c^{d+}_{\text{R2}}
\end{array}
\right)
=
\left(
\begin{array}{cccc}
 0 & 0 & e^{i \left(\chi_u-\Phi _u\right)} & 0 \\
 0 & 0 & 0 & e^{i \left(\chi_d-\Phi _d\right)} \\
 e^{i \left(\Phi _u+\chi_u\right)} & 0 & 0 & 0 \\
 0 & e^{i \left(\Phi _d+\chi_d\right)} & 0 & 0 \\
\end{array}
\right)_{S_R}
\left(
\begin{array}{c}
 c^{u+}_{\text{R1}} \\
 c^{d+}_{\text{R1}}\\
 c^{u-}_{\text{R2}} \\
 c^{d-}_{\text{R2}}
\end{array}
\right)
\end{equation}
where $\chi_i$ are the dynamical phases that electrons acquire while traveling
in each arm $i = u,d$.
Moreover, the application of the magnetic flux $\Phi$ across the AB ring causes the particles to acquire additional phases on each arm (namely, $\Phi_u$ in the upper and $\Phi_d$ in the downer arm), such that $\Phi_u-\Phi_d=2\pi\Phi/\Phi_0$ with $\Phi_0=h/e$ being the the flux quantum. 
As a result, right moving particles propagating in each arm ($i = u,d$) acquire a global dynamical phase $\chi_{i}+\Phi_{i}$, while left moving particle get a global phase $\chi_{i}-\Phi_{i}$.
Also in this case $S_R$ can be written in the form
\begin{equation}
 S_R= 
 \begin{pmatrix}
r_R & t_R^{\prime} \\ 
t_R &  r_R^{\prime}
\end{pmatrix}.
\end{equation}
As shown in Refs. \cite{Haack2019,Haack2021}, by linearizing the spectrum around the Fermi energy $\epsilon_F$ and taking into account an additional voltage gate $V_g$ applied to the lower arm of the ring, we can write the dynamical phases $\chi_u$ and $\chi_d$ in the following way
\begin{equation}
\label{chi_u_app}
\chi_u=\left[k_F+\frac{\epsilon-\epsilon_F}{\hbar v_F}\right]L_u\equiv\left[\frac{\xi}{\lambda_F}+\frac{\epsilon-\epsilon_F}{\Delta}\right]\frac{L_u}{\xi}
\end{equation}
\begin{equation}
\label{chi_d_app}
\chi_d=\left[k_F+\frac{\epsilon-(\epsilon_F+V_g)}{\hbar v_F}\right]L_d\equiv\left[\frac{\xi}{\lambda_F}+\frac{\epsilon-(\epsilon_F+V_g)}{\Delta}\right]\frac{L_d}{\xi}
\end{equation}
with $L_u$ and $L_d$ the lengths of the upper and downer arm respectively, and where we introduced the Fermi wave vector $k_F=\sqrt{2m\epsilon_F}/\hbar=1/\lambda_F$ (with $\lambda_F$ the Fermi wave length). In Eqs. \eqref{chi_u_app} and \eqref{chi_d_app}, we expressed all the quantities with respect to the superconducting energy gap $\Delta$ and the coherence length $\xi=\hbar v_F/\Delta$.

\subsection{Combination of the scattering matrices}

\label{sub_app_comb_scattering_matrix}
By following Refs.~\cite{Datta1997,Gresta2021} we first combine matrices $S_{T_1}$ of Eq.~(\ref{S1}) and $S_R$ of Eq.~(\ref{SR}), and obtain
\begin{equation}
    S_{T_1} \circ S_R = \begin{pmatrix}
r & t' \\ 
t &  r'
\end{pmatrix},
\end{equation}
in which
\begin{align}
&r=r_1+t'_1r_R\left[\mathbb{1}-r'_1r_R \right ]^{-1}t_1\nonumber\\
&r'=r'_R+t_R\left[\mathbb{1}-r'_1r_R \right ] ^{-1}r'_1t'_R\nonumber\\
&t= t_R\left[\mathbb{1}-r'_1r_R \right ]^{-1}t_1\nonumber\\
&t'=t'_1\left[\mathbb{1}-r_Rr'_1 \right ]^{-1}t'_R,
\end{align}
where $\mathbb{1}$ stands for the $2\times 2$ identity matrix.
Finally, by applying the same procedure but adding $S_{T_2}$ we obtain the particle-scattering matrix of the AB ring
\begin{equation}
    S_{AB}^{e} \equiv S_{T_1}\circ S_R \circ S_{T_2}
\end{equation}
which takes the following form
\begin{align}
\label{SAB}
S_{AB}^{e}&= \begin{pmatrix}
r_{AB}^{ee} & t_{AB}^{\prime ee} \\ 
t_{AB}^{ee} &  r_{AB}^{\prime ee}
\end{pmatrix}\nonumber\\&
=\left(
\begin{array}{cc}
 \frac{f_- \cos (\delta\chi)-f_+ \cos (\Phi )+4 \sqrt{1-\tau} \cos (\chi)}{f_\pm \cos (\delta\chi)+f_+ \cos (\Phi )+2 i \tau \sin (\chi)+2 (\tau-2) \cos (\chi)} & \frac{2 i \tau e^{-i \Phi/2 } \left(\sin \left(\chi_u\right)+e^{i \Phi } \sin \left(\chi_d\right)\right)}{f_- \cos (\delta \chi)+f_+ \cos (\Phi )+2 i \tau \sin (\chi)+2 (\tau-2) \cos (\chi)} \\
 \frac{2 i \tau e^{-i \Phi/2 } \left(\sin \left(\chi_d\right)+e^{i \Phi } \sin \left(\chi_u\right)\right)}{f_- \cos (\delta \chi)+f_+ \cos (\Phi )+2 i \tau \sin (\chi)+2 (\tau-2) \cos (\chi)}
   & \frac{f_- \cos (\delta \chi)-f_+ \cos (\Phi )+4 \sqrt{1-\tau} \cos (\chi)}{f_- \cos (\delta \chi)+f_+ \cos (\Phi )+2 i \tau \sin (\chi)+2 (\tau-2) \cos (\chi)} \\
\end{array}
\right)
\end{align}
in which the upper index $e$ indicates that such scattering matrix only relates incoming with outgoing particles (namely electrons), and where we defined the quantities: $\chi=\chi_u+\chi_d$, $\delta \chi=\chi_u-\chi_d$, $f_-=(\sqrt{1-\tau}-1)^2$ and $f_+=(\sqrt{1-\tau}+1)^2$.
From Eq. \ref{SAB} we can compute the transmission function which is given by
\begin{equation}
\label{TAB_eq}
T_{AB}=\abs{t_{AB}^{ee}}^2=\abs{t_{AB}^{\prime ee}}^2=\frac{4 \tau^2 \left(2 \cos (\Phi ) \sin \left(\chi_d\right) \sin \left(\chi_u\right)+\sin ^2\left(\chi_d\right)+\sin ^2\left(\chi_u\right)\right)}{\left[f_-
   \cos (\delta\chi)+f_+ \cos (\Phi )+2 (\tau-2) \cos (\chi)\right]^2+4 \tau^2 \sin^2(\chi)}.
\end{equation}

\subsection{Extension with the BdG formalism}
\label{sub_app_presence_superconductor_A3}

In order to properly describe the transport in presence of a superconducting component, it is mandatory to take into account for holes. In this respect the particle-scattering matrix $S_{AB}^{e}$ of Eq.~\eqref{SAB}, can be extended in the BdG formalism by changing the sign of the energy and taking the complex conjugate as specified by the particle-hole symmetry relations already introduced in Eq.~\eqref{particle_hole_transformations}. As a result we obtain the expression of the $S_{AB}$ full scattering matrix as presented in Eq.~\eqref{S_AB} of the main text, which we propose here again making explicit the scattering basis:
\begin{equation}
\label{SAB_super}
\left(
\begin{array}{c}
 c^-_1 \\
 b^-_1 \\
 c^+_2 \\
 b^+_2 \\
\end{array}
\right)
=
\left(
\begin{array}{cccc}
 r^{ee}_{\text{AB}}(\epsilon ) & 0 & t'^{ee}_{\text{AB}}(\epsilon ) & 0 \\
 0 & r^{hh }_{\text{AB}}(-\epsilon )^* & 0 & t'^{hh}_{\text{AB}}(-\epsilon )^* \\
 t^{ee }_{\text{AB}}(\epsilon ) & 0 & r'^{ee}_{\text{AB}}(\epsilon ) & 0 \\
 0 & t^{hh }_{\text{AB}}(-\epsilon )^* & 0 & r'^{hh}_{\text{AB}}(-\epsilon )^* \\
\end{array}
\right)_{S_{AB}}
\left(
\begin{array}{c}
 c^+_1 \\
 b^+_1 \\
 c^-_2 \\
 b^-_2 \\
\end{array}
\right)
\end{equation}
where we indicated with $b_i^{\pm}$ the incoming and outgoing holes in the left ($i=1$) and the right ($i=2$) lead respectively. 
As already mentioned in the main text, in Eq.~\eqref{SAB_super}, each submatrix takes a block-diagonal form since in the ring an electron cannot be converted into a hole or vice versa.

\section{N-S junction scattering matrix}
\label{appendix_NS}

The scattering matrix equation for $S_{NS}$, describing the N-S interface between the ring and the superconducting lead, can be written as
\begin{equation}
\label{SSC}
\left(
\begin{array}{c}
 c^-_2 \\
 b^-_2 \\
 c_{S}^+ \\
 b_{S}^+ \\
\end{array}
\right)
=
\left(
\begin{array}{cccc}
 r_{ee} & r_{eh} & t_{e \tilde{e}} & t_{e \tilde{h}} \\
 r_{he} & r_{hh} & t_{h \tilde{e}} & t_{h \tilde{h}} \\
 t_{e \tilde{e}} & t_{h \tilde{e}} & r_{\tilde{e}\tilde{e}} & r_{\tilde{e} \tilde{h}} \\
 t_{e \tilde{h}} & t_{h \tilde{h}} & r_{\tilde{e}\tilde{h}} & r_{\tilde{h}\tilde{h}} \\
\end{array}
\right)_{S_{NS}}
\left(
\begin{array}{c}
 c^+_2 \\
 b^+_2 \\
 c_{S}^- \\
 b_{S}^- \\
\end{array}
\right)
\end{equation}
where we indicate with $c_S^{\pm}/b_S^{\pm}$ the incoming and outgoing quasiparticles/quasiholes in the superconductor, with $\pm$ labeling the direction of propagation of quasiparticles ($+$ for right movers and $-$ for left movers).
In Eq. \ref{SSC}, the scattering coefficients have been obtained by solving the wave function matching problem at the interface between the ring and the superconducting contact in the so-called Andreev approximation limit (when $\epsilon,\Delta\ll\epsilon_F$), and take the following form
\begin{subequations}
\label{scattering_coeff_Andreev_approx}
\begin{align}
\label{ree}
r_{ee}&=-\frac{Z\left(i+Z\right)\left(u_0^2-v_0^2\right)}{u_0^2+Z^2\left(u_0^2-v_0^2\right)}\\
\label{rhe}
r_{he}&=\frac{u_0v_0}{u_0^2+Z^2\left(u_0^2-v_0^2\right)}e^{-i\phi}\\
\label{tEe}
t_{\tilde{e}e}&=\frac{\left(1-iZ\right)u_0\sqrt{u_0^2-v_0^2}}{u_0^2+Z^2\left(u_0^2-v_0^2\right)}e^{-i\frac{\phi}{2}}\cdot \Theta(\epsilon-\Delta)\\
\label{tHe}
t_{\tilde{h}e}&=\frac{Z v_0\sqrt{u_0^2-v_0^2}}{u_0^2+Z^2\left(u_0^2-v_0^2\right)}e^{-i\frac{\phi}{2}}\cdot \Theta(\epsilon-\Delta)
\end{align}
\end{subequations}
while the remaining scattering coefficients respect the following relations
\begin{equation}
\begin{dcases}
 r_{\tilde{e}\tilde{e}}=r_{\tilde{h}\tilde{h}}^*=r_{hh}^*=r_{ee}\\
-r_{\tilde{h}\tilde{e}}e^{-i\phi}=-r_{\tilde{e}\tilde{h}}^*e^{-i\phi}=r_{eh}^*=r_{he}\\
t_{e\tilde{e}}e^{-i\phi}=t_{h\tilde{h}}^*e^{-i\phi}=t_{\tilde{h}h}^*=t_{\tilde{e}e}\\
-t_{h\tilde{e}}=-t_{e\tilde{h}}^*=t_{\tilde{e}h}^*=t_{\tilde{h}e}
\end{dcases}.
\end{equation}
As already mentioned in the main text, it is important to notice that, within the Andreev approximation, all the details about the curvature of the eigenspectrum dispersion relations are completely lost. As consequence, themoelectric effects may result strongly suppressed.  
Instead, for a better description of the thermoelectric phenomena in hybrid superconducting systems, it is necessary to go beyond the Andreev approximation (see \ref{appendix_C} for more details).
In Eqs. \ref{scattering_coeff_Andreev_approx} we introduced the so-call coherence factors $u_0$ and $v_0$ which take the following form
\begin{align}
u_0(\epsilon)&=\sqrt{\frac{1}{2}\left(1+\sqrt{\frac{\epsilon^2-\Delta^2}{\epsilon^2}}\right)}\equiv\sqrt{\frac{\Delta}{2\epsilon}}e^{\frac{1}{2}h(\epsilon)}\nonumber\\ v_0(\epsilon)&=\sqrt{\frac{1}{2}\left(1-\sqrt{\frac{\epsilon^2-\Delta^2}{\epsilon^2}}\right)}\equiv\sqrt{\frac{\Delta}{2\epsilon}}e^{-\frac{1}{2}h(\epsilon)}
\end{align}
with
\begin{equation}
\label{h}
h(\epsilon)\equiv\begin{dcases}
 \arccosh{\left(\frac{\epsilon}{\Delta}\right)}&\text{for}\quad \epsilon>\Delta\\
 i \arccos{\left(\frac{\epsilon}{\Delta}\right)}&\text{for}\quad \epsilon<\Delta
\end{dcases}.
\end{equation}
Moreover, the scattering coefficients of Eqs.~\eqref{scattering_coeff_Andreev_approx}, depend explicitly on the dimensionless transparency parameter $Z$ which characterizes the interface with the superconductor~\cite{BTK1982}.
%

\section{Beyond the Andreev approximation}
\label{appendix_C}

Here we present the explicit expressions of the scattering coefficients of Eq.~\eqref{S_NS}, obtained beyond the Andreev approximation in the case of an ideal interface $Z=0$: 
{\small
\begin{align*}
\begin{split}
r_{ee}=&\frac{e^{2 i x k_e} \left(-\Gamma  q_{\tilde{e}} q_{\tilde{h}}+\Gamma  k_e k_h+\Xi _1 k_e-\Xi _2 k_h\right)}{\Gamma  q_{\tilde{e}}
   q_{\tilde{h}}+\Gamma  k_e k_h+\Xi _1 k_e+\Xi _2 k_h}\\
   r_{he}=&\frac{2 u_0 v_0 \sqrt{k_e k_h} \left(q_{\tilde{e}}+q_{\tilde{h}}\right) e^{i \left(x
   k_e-x k_h-\phi \right)}}{\Gamma  q_{\tilde{e}} q_{\tilde{h}}+\Gamma  k_e k_h+\Xi _1 k_e+\Xi _2 k_h}\\
   t_{\tilde{e}e}=&\frac{2 u_0
   \left(q_{\tilde{h}}+k_h\right) \sqrt{\Gamma  k_e q_{\tilde{e}}} e^{\frac{1}{2} i \left(-2 x q_{\tilde{e}}+2 x k_e-\phi \right)}}{\Gamma 
   q_{\tilde{e}} q_{\tilde{h}}+\Gamma  k_e k_h+\Xi _1 k_e+\Xi _2 k_h}\\
   t_{\tilde{h}e}=&-\frac{2 v_0 \left(k_h-q_{\tilde{e}}\right) \sqrt{\Gamma  k_e
   q_{\tilde{h}}} e^{\frac{1}{2} i \left(2 x q_{\tilde{h}}+2 x k_e-\phi \right)}}{\Gamma  q_{\tilde{e}} q_{\tilde{h}}+\Gamma  k_e k_h+\Xi _1
   k_e+\Xi _2 k_h}
\end{split}
\qquad
\begin{split}
r_{eh}=&\frac{2 u_0 v_0 \sqrt{k_e k_h} \left(q_{\tilde{e}}+q_{\tilde{h}}\right) e^{i \left(x k_e-x k_h+\phi \right)}}{\Gamma  q_{\tilde{e}}
   q_{\tilde{h}}+\Gamma  k_e k_h+\Xi _1 k_e+\Xi _2 k_h}\\
   r_{hh}=&\frac{e^{-2 i x k_h} \left(-\Gamma  q_{\tilde{e}} q_{\tilde{h}}+\Gamma  k_e k_h-\Xi _1
   k_e+\Xi _2 k_h\right)}{\Gamma  q_{\tilde{e}} q_{\tilde{h}}+\Gamma  k_e k_h+\Xi _1 k_e+\Xi _2 k_h}\\
   t_{\tilde{e}h}=&-\frac{2 v_0
   \left(k_e-q_{\tilde{h}}\right) \sqrt{\Gamma  k_h q_{\tilde{e}}} e^{-\frac{1}{2} i \left(2 x q_{\tilde{e}}+2 x k_h-\phi \right)}}{\Gamma 
   q_{\tilde{e}} q_{\tilde{h}}+\Gamma  k_e k_h+\Xi _1 k_e+\Xi _2 k_h}\\
   t_{\tilde{h}h}=&\frac{2 u_0 \left(q_{\tilde{e}}+k_e\right) \sqrt{\Gamma  k_h
   q_{\tilde{h}}} e^{i x q_{\tilde{h}}-i x k_h+\frac{i \phi }{2}}}{\Gamma  q_{\tilde{e}} q_{\tilde{h}}+\Gamma  k_e k_h+\Xi _1 k_e+\Xi _2
   k_h}
\end{split}
\end{align*}
}

{\small
\begin{align}
\begin{split}
r_{\tilde{e}\tilde{e}}=&\frac{e^{-2 i x q_{\tilde{e}}} \left(\Gamma  q_{\tilde{e}} q_{\tilde{h}}-\Gamma  k_e k_h-\Xi_3 k_e+\Xi _4 k_h\right)}{\Gamma 
   q_{\tilde{e}} q_{\tilde{h}}+\Gamma  k_e k_h+\Xi _1 k_e+\Xi_2 k_h}\nonumber\\
   r_{\tilde{h}\tilde{e}}=&-\frac{2 u_0 v_0 \left(k_e+k_h\right) \sqrt{q_{\tilde{e}} q_{\tilde{h}}}
   e^{-i x \left(q_{\tilde{e}}-q_{\tilde{h}}\right)}}{\Gamma  q_{\tilde{e}} q_{\tilde{h}}+\Gamma  k_e k_h+\Xi _1 k_e+\Xi _2 k_h}\nonumber\\
   t_{e\tilde{e}}=&\frac{2 u_0
   \left(q_{\tilde{h}}+k_h\right) \sqrt{\Gamma  k_e q_{\tilde{e}}} e^{\frac{1}{2} i \left(-2 x q_{\tilde{e}}+2 x k_e+\phi \right)}}{\Gamma 
   q_{\tilde{e}} q_{\tilde{h}}+\Gamma  k_e k_h+\Xi _1 k_e+\Xi _2 k_h}\nonumber\\
   t_{h\tilde{e}}=&-\frac{2 v_0 \left(k_e-q_{\tilde{h}}\right) \sqrt{\Gamma  k_h
   q_{\tilde{e}}} e^{-\frac{1}{2} i \left(2 x q_{\tilde{e}}+2 x k_h+\phi \right)}}{\Gamma  q_{\tilde{e}} q_{\tilde{h}}+\Gamma  k_e k_h+\Xi _1
   k_e+\Xi _2 k_h}
\end{split}
\qquad
\begin{split}
r_{\tilde{e}\tilde{h}}=&-\frac{2 u_0 v_0 \left(k_e+k_h\right) \sqrt{q_{\tilde{e}} q_{\tilde{h}}} e^{-i x \left(q_{\tilde{e}}-q_{\tilde{h}}\right)}}{\Gamma 
   q_{\tilde{e}} q_{\tilde{h}}+\Gamma  k_e k_h+\Xi _1 k_e+\Xi _2 k_h}\\
   r_{\tilde{h}\tilde{h}}=&\frac{e^{2 i x q_{\tilde{h}}} \left(\Gamma  q_{\tilde{e}}
   q_{\tilde{h}}-\Gamma  k_e k_h+\Xi _3 k_e-\Xi _4 k_h\right)}{\Gamma  q_{\tilde{e}} q_{\tilde{h}}+\Gamma  k_e k_h+\Xi _1 k_e+\Xi _2
   k_h}\\
   t_{e\tilde{h}}=&-\frac{2 v_0 \left(k_h-q_{\tilde{e}}\right) \sqrt{\Gamma  k_e q_{\tilde{h}}} e^{\frac{1}{2} i \left(2 x q_{\tilde{h}}+2 x k_e+\phi
   \right)}}{\Gamma  q_{\tilde{e}} q_{\tilde{h}}+\Gamma  k_e k_h+\Xi _1 k_e+\Xi _2 k_h}\\
   t_{h\tilde{h}}=&\frac{2 u_0 \left(q_{\tilde{e}}+k_e\right)
   \sqrt{\Gamma  k_h q_{\tilde{h}}} e^{-\frac{1}{2} i \left(-2 x q_{\tilde{h}}+2 x k_h+\phi \right)}}{\Gamma  q_{\tilde{e}}
   q_{\tilde{h}}+\Gamma  k_e k_h+\Xi _1 k_e+\Xi _2 k_h}
\end{split}
\end{align}
}
where we defined the quantities
\begin{equation}
\Gamma=u_0^2-v_0^2\nonumber,
\end{equation}
and
\begin{equation}
\mqty(\Xi_1&\Xi_2\\\Xi_3&\Xi_4)=\mqty(q_{\tilde{h}}&q_{\tilde{e}}\\q_{\tilde{h}}&q_{\tilde{e}})u_0^2+\mqty(q_{\tilde{e}}&q_{\tilde{h}}\\-q_{\tilde{e}}&-q_{\tilde{h}})v_0^2\nonumber.
\end{equation}
Above we introduced the electron and hole wave vector amplitudes
\begin{align}
k_e=k_F\sqrt{1+\frac{\epsilon}{\epsilon_F}}; && k_h=k_F\sqrt{1-\frac{\epsilon}{\epsilon_F}}
\end{align}
with $k_F=\sqrt{2m\epsilon_F}/\hbar$ the Fermi wave vector, and the wave vector amplitudes 
\begin{align}
\label{qe_qh}
q_{\tilde{e}}=k_F\sqrt{1+\sqrt{\frac{\epsilon^2-\Delta^2}{\epsilon_F^2}}}; && q_{\tilde{h}}=k_F\sqrt{1-\sqrt{\frac{\epsilon^2-\Delta^2}{\epsilon_F^2}}} 
\end{align}
for quasiparticles and quasiholes respectively.\\

\nocite{*}
\bibliographystyle{iopart-num}
\bibliography{AB_Ring_biblio}

\end{document}